\documentclass[12pt]{amsart}
\usepackage{latexsym}
\usepackage[dvips]{graphicx}
\usepackage{amsmath,amssymb, amsthm}
\usepackage{graphicx, color}

\newtheorem{thm}{Theorem}

\newtheorem{lem}[thm]{Lemma}
\theoremstyle{definition}

\newtheorem{remark}[thm]{Remark}

\newcommand{\bA}{\mathsf{A}}
\newcommand{\bB}{\mathsf{B}}

\newcommand{\bE}{\mathrm{E}}

\newcommand{\Dt}{\Delta}

\newcommand{\im}{\text{i}}

\newcommand{\al}{\alpha}
\newcommand{\de}{\delta}

\title{Approximation of eigenvalues of spot cross volatility matrix with a view
toward principal component analysis
}

\author{Nien-Lin Liu}
\author{Hoang-Long Ngo}
\address[Nien-Lin Liu and Hoang-Long Ngo]{Research Organization of Science and Engineering, Ritsumeikan University\\
1-1-1 Nojihigashi - Kusatsu - Shiga - Japan}
\address[Hoang-Long Ngo]{Japan Science and Technology Agency}
\address[Hoang-Long Ngo]{Hanoi National University of Education\\
136 Xuan Thuy - Cau Giay - Hanoi - Vietnam}

\date{}
\numberwithin{equation}{section}
\begin{document}
\maketitle
\thispagestyle{empty}
\begin{abstract}
In order to study the geometry of interest rates market dynamics,  Malliavin, Mancino and  Recchioni \emph{ [A non-parametric calibration of the HJM geometry: an application of It\^o calculus to financial statistics, {\it Japanese Journal of  Mathematics}, 2, pp.55--77, 2007]} introduced a scheme, which is based on the Fourier Series method, to estimate eigenvalues of a spot cross volatility matrix. In this paper, we present another estimation scheme based on the Quadratic Variation method. We first establish limit theorems for each scheme and then we use a stochastic volatility model of Heston's type to compare the effectiveness of these two schemes.
\end{abstract}

%%%%%%%%%%%%%%%%%%%%%%%%%%%%%%%%%%%%%%%%%%%%%%%%%%
%%%%%%%%%%%%%%%%%%%%%%%%%%%%%%%%%%%%%%%%%%%%%%%%%%
\section{Introduction}
Let $X$ be a $d$-dimensional stochastic process defined on a probability space 
$(\Omega, \mathcal{F}, (\mathcal{F})_t, P)$ by
\begin{equation} \label{eqX}
d X (t) = \bA (t, w)dt + \bB (t,w)dW(t), \ 0 \leq  t \leq T, 
\end{equation}
where $W$ is a $d_1$-dimensional standard Brownian motion, $\bA$ is a $d$-dimensional drift process and $\bB$ is a $R^{d\times d_1}$-valued c\`adl\`ag volatility process. In mathematical finance it is widely accepted that processes $X$ of the form defined by (\ref{eqX}) are reasonable models for the (log return of) price processes and interest rates.

The spot cross volatility matrix $\Sigma = (\Sigma_{i,j})_{1\leq i, j \leq d}$ of process $X$ is defined by
$$\Sigma_{i,j}(t)  = \sum_{k=1}^{d_1} \bB_{i,k}(t)\bB_{j,k}(t), \quad 0 \leq t \leq T.$$

We are interested in the following problem: given a finite set of observation data $\{X(t_k, \omega_0): \ t_k = kT/n, k = 0,\ldots, n\}$ of a single trajectory $\omega_0 \in \Omega$, we want to estimate the eigenvalues of $\Sigma(t, \omega_0)$ for any $t \in [0.T]$. This problem appears in mathematical finance, especially in principal component analysis (see \cite{AL2011, L2010, MMR2007}). The estimation of the eigenvalues of the integrated volatility matrix was studied by Wang and Zou (\cite{WZ2010}) (see also the references therein). By the time we completed this paper, we learnt that Jacod and Podolskij \cite{JP2012} had previously introduced some statistics based on a random perturbation approach for ranks of volatility metric of continuous It\^o process. Our approach differs from that in \cite{JP2012} and can be applied to It\^o processes with jump components.

Our method to solve this problem is as follows: first we approximate the spot cross volatility matrix $\Sigma$ by a matrix $\hat{\Sigma}$ using the given observations of $X$; next we approximate the eigenvalues of $\Sigma$ by those of $\hat{\Sigma}$. 

The spot volatility estimation is an important problem in mathematical finance and has been extensively studied by many authors. Up to now, there are two main approaches to this problem. The first approach called the Fourier Series method was introduced by  \cite{MM2002} and later developed in \cite{MMR2007, MM2009}. The second approach called the Quadratic Variation method was introduced by  \cite{OW2007} and later developed in \cite{O2008, ON2010, NO2009} (see also \cite{APPS2012, JR2012}). It should be noted that there is a very rich literature on the problem of measuring the so called realized volatility as well as problem of estimating parameters of diffusion processes, see \cite{Rao1999, BS2004, AsH2010, BHL2012} and the references therein.

In this paper, we present some limit theorems and a numerical study to analyze the effectiveness of the estimation of eigenvalues by using Fourier Series and Quadratic Variation methods. It should be mentioned that in reality, one cannot observe directly either cross volatility matrix or the eigenvalues. Therefore we perform a numerical study with dummy data for which we know both the volatility matrix and its eigenvalues beforehand. In particular, we show that the Fourier Series method may lead to some unexpected results when estimating small eigenvalues; this situation would never arise using the Quadratic Variation method.

\subsection*{Acknowledgment}  The authors thank Jiro Akahori, Freddy Delbaen, Arturo Kohatsu-Higa, Maria Elvira Mancino, and Shigeyoshi Ogawa for their helpful comments. The authors are also grateful to the referee for her/his valuable comments which led to improvement of the paper.

%%%%%%%%%%%%%%%%%%%%%%%%%%%%%%%%%%%%%%%%%%%%%%%%%%
%%%%%%%%%%%%%%%%%%%%%%%%%%%%%%%%%%%%%%%%%%%%%%%%%%
\section{The first Fourier Series estimation scheme}\label{FS}
In a series of papers \cite{MM2002, MMR2007, MM2009}, Malliavin et al. introduced a number of Fourier Series estimation schemes for spot volatilities. Although these schemes are essentially based on a same idea, they are slightly different. As we will present later, each scheme has both advantages and disadvantages compared to the other. 

In this section, we summarize the Fourier Series method presented in \cite{MM2002}. By a change of origin and rescaling, one can suppose that $T = 2\pi$ and the Fourier Series method reconstructs $\Sigma_.(t)$ for all $t \in (0,2\pi)$. Let us denote the Fourier coefficients of $dX_j, \ j = 1,\ldots, d,$ by
\begin{align*}
a_k(dX_j) = \frac{1}{\pi} \int_{(0,2\pi)}\cos(kt)dX_j(t),
b_k(dX_j) = \frac{1}{\pi} \int_{(0,2\pi)}\sin(kt)dX_j(t).
\end{align*}
The Fourier coefficients of each cross volatility $\Sigma_{u,v},\ 1 \leq u, v \leq d,$ are defined by
\begin{align*}
a_k(\Sigma_{u,v}) = \frac{1}{\pi} \int_{(0,2\pi)}\cos(kt)\Sigma_{u,v}(t)dt,
b_k(\Sigma_{u,v}) = \frac{1}{\pi}\int_{(0,2\pi)} \sin(kt)\Sigma_{u,v}(t)dt.
\end{align*}
It follows from the Fourier-F\'ejer inversion formula that one can reconstruct $\Sigma$ from its Fourier coefficients by
\begin{align*}
\Sigma_{u,v}(t) = \lim_{N \to \infty} \sum_{k=0}^N \big( 1 - \frac{k}{N}\big) \big(a_k(\Sigma_{u,v})\cos(kt) + b_k(\Sigma_{u,v})\sin(kt) \big).
\end{align*} 
In practice, based on the observation of $X$ at times $t_i = 2\pi i/n, \ i = 0,\ldots, n$, one can approximate $\Sigma$ as follows. We fix some positive integer $N$.
\begin{enumerate}
\item Fourier coefficients $a_k(dX_j), \ b_k(dX_j), \ k = 0,\ldots, 2N,$ are approximated by
\begin{align*}
\hat{a}_k(dX_j) &= \frac{1}{\pi} \sum_{i=1}^n \big(\cos( kt_{i-1}) - \cos(kt_i) \big)X_j(t_{i-1}) + \frac{1}{\pi} \big( X_j(t_n) - X_j(t_0) \big),\\
\hat{b}_k(dX_j)  &= \frac{1}{\pi} \sum_{i=1}^n \big(\sin( kt_{i-1}) - \sin(kt_i) \big)X_j(t_{i-1}).
\end{align*}
\item Fourier coefficients of each cross volatility $\Sigma_{u,v}, \ 1 \leq u, v \leq d,$ are approximated by
\begin{align*}
\hat{a}_0(\Sigma_{u,v}) & = \frac{\pi}{2(N + 1 - n_0)} \sum_{s = n_0}^{N} \big( \hat{a}_s(dX_u) \hat{a}_s(dX_v) + \hat{b}_s(dX_u) \hat{b}_s(dX_v)\big),\\
\hat{a}_k(\Sigma_{u,v}) &= \frac{\pi}{N + 1 - n_0} \sum_{s = n_0}^{N} \big( \hat{a}_s(dX_u) \hat{a}_{s+k}(dX_v) + \hat{a}_s(dX_v) \hat{a}_{s+k}(dX_u)\big), \\
\hat{b}_k(\Sigma_{u,v}) &= \frac{\pi}{N + 1 - n_0} \sum_{s = n_0}^{N} \big( \hat{a}_s(dX_u) \hat{b}_{s+k}(dX_v) + \hat{a}_s(dX_v) \hat{b}_{s+k}(dX_u)\big),
\end{align*}
for each $k = 0,\ldots, N$.
\item The volatilities $\Sigma_{u,v}(t)$ are approximated by 
\begin{equation} \label{def_hatSm}
\hat{\Sigma}^{N,n}_{u,v}(t) = \sum_{k=0}^N \big( 1 - \frac{k}{N}\big)\big(\hat{a}_k(\Sigma_{u,v}))\cos(kt) + \hat{b}_k(\Sigma_{u,v})\sin(kt) \big).
\end{equation} 
\end{enumerate}
Sometime, it is preferable to smooth the F\'ejer kernel in (\ref{def_hatSm}) by replacing $(1-k/N)$ with $\sin^2(\delta k)/(\delta k)^2$ for some appropriate parameter $\delta >0$.

\begin{remark}
It should be noted here that although the matrix $\hat{\Sigma}^{N,n}(t)$ is symmetric, it is not non-negative definite in general. Therefore some of its eigenvalues may be negative, which is not expected in practice.
\end{remark}

%%%%%%%%%%%%%%%%%%%%%%%%%%%%%%%%%%%%%%%%%%%%%%%%%%
%%%%%%%%%%%%%%%%%%%%%%%%%%%%%%%%%%%%%%%%%%%%%%%%%%
\section{The second Fourier Series estimation scheme}\label{FS2}
In \cite{MM2009}, the authors introduced another version of Fourier Series estimation scheme. Their new scheme was designed to deal with asynchronous data. In the following, we will specialize it for the case of regular sampling. 
We define
$\de^j_i := X^{j}(t_{i+1}) - X^{j}(t_{i}), \quad j = 1, \ldots, d.$
For any integer $k$, $|k| \leq 2N$, let
\begin{align*}
c^j_{k} := \frac{1}{2\pi}\sum^{n-1}_{i=0} e^(-\im kt_{i})\de^j_i, \quad j = 1, \ldots, d.
\end{align*}
For each $1 \leq j_1 \leq j_2 \leq d,$
let $\al_{k}(N, j_1, j_2)$ for $|k| \leq N$ be given by
\begin{align*}
\al_{k}(N, j_1, j_2) := \frac{2\pi}{2N+1}\sum_{|s| \leq N} c^{j_1}_{s}c^{j_2}_{k-s}.
\end{align*}
Finally, define
\begin{align*}
\Sigma^{j_1 j_2}_{n, N}(t) &:= \sum_{|k| \leq N} \Big(1-\frac{|k|}{N}\Big)\al_{k}(N, j_1, j_2)e^(\im kt)\\
&= \al_{0}(N, j_1, j_2) + \sum^{N}_{k=1} \Big(1-\frac{k}{N}\Big)\Big(\al_{k}(N, j_1, j_2)e^{\im kt}+\al_{-k}(N, j_1, j_2)e^{-\im kt}\Big).
\end{align*}
Since the above estimator is written with complex numbers, it may be inconvenient to do simulation. Therefore we rewrite it as below:
\begin{equation} 
\Sigma^{j_1 j_2}_{n, N}(t) := \al_{0}(N, j_1, j_2) + \sum_{k=1}^N \Big(1 -\frac{k}{N}\Big)\Big(a^{j_1 j_2}_k\cos(kt)+b^{j_1 j_2}_k\sin(kt)\Big),
\label{Gammaold}
\end{equation}
where 
\begin{align*}
a^{j_1 j_2}_k &= \frac{1}{\pi(2N+1)}\Big[ \sum^{N}_{s=1} \Big\{\hat{a}_{s}(dX^{j_1})\hat{a}_{k-s}(dX^{j_2}) - \hat{b}_{s}(dX^{j_1})\hat{b}_{k-s}(dX^{j_2})\\
&\hskip 1cm + \hat{a}_{s}(dX^{j_1})\hat{a}_{k+s}(dX^{j_2}) + \hat{b}_{s}(dX^{j_1})\hat{b}_{k+s}(dX^{j_2})\Big\} + \hat{a}_{k}(dX^{j_2})\Big(X^{j_1}(2\pi) - X^{j_1}(0)\Big)\Big],\\
b^{j_1 j_2}_k &= \frac{1}{\pi(2N+1)}\Big[ \sum^{N}_{s=1} \Big\{\hat{a}_{s}(dX^{j_1})\hat{b}_{k-s}(dX^{j_2}) + \hat{b}_{s}(dX^{j_1})\hat{a}_{k-s}(dX^{j_2})\\
&\hskip 1cm + \hat{a}_{s}(dX^{j_1})\hat{b}_{k+s}(dX^{j_2}) - \hat{b}_{s}(dX^{j_1})\hat{a}_{k+s}(dX^{j_2})\Big\} + \hat{b}_{k}(dX^{j_2})\Big(X^{j_1}(2\pi) - X^{j_1}(0)\Big)\Big],
\end{align*}
and,
\begin{align*}
&\hat{a}_{s}(dX^{j_1}) = \sum_{i}\cos(st_{i})\de^{j_1}_{i},
\quad \hat{a}_{s}(dX^{j_2}) = \sum_{j}\cos(st_{j})\de^{j_2}_{j},\\
&\hat{b}_{s}(dX^{j_1}) = \sum_{i}\sin(st_{i})\de^{j_1}_{i},
\quad \hat{b}_{s}(dX^{j_2}) = \sum_{j}\sin(st_{j})\de^{j_2}_{j},
\end{align*}
and
\begin{align*}
\al_{0}(N, j_1, j_2)
&= \frac{1}{2\pi(2N+1)}\Big[\sum^{N}_{s=1} 2\Big\{\hat{a}_{s}(dX^{j_1})\hat{a}_{s}(dX^{j_2}) + \hat{b}_{s}(dX^{j_1})\hat{b}_{s}(dX^{j_2})\Big\}\\
&\hskip 4cm + \Big(X^{j_2}(2\pi) - X^{j_2}(0)\Big)\Big(X^{j_1}(2\pi) - X^{j_1}(0)\Big)\Big].
\end{align*}

\begin{remark}
Since the matrix $\Sigma^{j_1 j_2}_{n, N}(t)$ is not symmetric, the eigenvalues may not be real numbers. In order to overcome this drawback we propose two symmetrization methods as follows.
\end{remark}

%%%%%%%%%%%%%%%%%%%%%%%%%%%%%%%%%%%%%%%%%%%%%%%%%%
%%%%%%%%%%%%%%%%%%%%%%%%%%%%%%%%%%%%%%%%%%%%%%%%%%
\subsection{The first symmetrization}
A naive idea to symmetrize the covariance matric is that one first calculates  $\Sigma^{j_1 j_2}_{n, N}$ using formula (\ref{Gammaold}) for all $1\leq j_1 \leq j_2 \leq d,$ and then puts $\Sigma^{j_2 j_1}_{n, N}:= \Sigma^{j_1 j_2}_{n, N}$.

\subsection{The second symmetrization}
Another way to symmetrize the covariance matric is as follows: 
Denote
\begin{align*}
\al_{k}(N, j_1, j_2) := \frac{\pi}{2N+1}\sum_{|s| \leq N} \big(c^{j_1}_{s}c^{j_2}_{k-s} + c^{j_2}_{s}c^{j_1}_{k-s}\big),
\end{align*}
for any $|k| \leq N$ and define
\begin{align*}
\Sigma^{j_1 j_2}_{n, N}(t) &:= \sum_{|k| \leq N} \Big(1-\frac{|k|}{N}\Big)\al_{k}(N, j_1, j_2)e^(\im kt)\\
&= \al_{0}(N, j_1, j_2) + \sum^{N}_{k=1} \Big(1-\frac{k}{N}\Big)\Big(\al_{k}(N, j_1, j_2)e^{\im kt}+\al_{-k}(N, j_1, j_2)e^{-\im kt}\Big).
\end{align*}
To simplify the simulation, we rewrite $\Sigma^{j_1 j_2}_{n, N}$ as follows
\begin{align*}
\Sigma^{j_1 j_2}_{n, N}(t) := \al_{0}(N, j_1, j_2) + \sum_{k=1}^N \Big(1 -\frac{k}{N}\Big)\Big(a^{j_1 j_2}_k\cos(kt)+b^{j_1 j_2}_k\sin(kt)\Big),
\end{align*}
where
\begin{align*}
a^{j_1 j_2}_k &= \frac{1}{2\pi(2N+1)}\Big[\sum^{N}_{s=1} \Big\{\hat{a}_{s}(dX^{1})\hat{a}_{k-s}(dX^{2}) - \hat{b}_{s}(dX^{1})\hat{b}_{k-s}(dX^{2}) + \hat{a}_{s}(dX^{1})\hat{a}_{k+s}(dX^{2})\\
&\quad + \hat{b}_{s}(dX^{1})\hat{b}_{k+s}(dX^{2}) + \hat{a}_{s}(dX^{2})\hat{a}_{k-s}(dX^{1}) - \hat{b}_{s}(dX^{2})\hat{b}_{k-s}(dX^{1}) + \hat{a}_{s}(dX^{2})\hat{a}_{k+s}(dX^{1})\\
&\quad + \hat{b}_{s}(dX^{2})\hat{b}_{k+s}(dX^{1})\Big\} + \hat{a}_{k}(dX^{2})\Big(X^{1}(2\pi) - X^{1}(0)\Big) + \hat{a}_{k}(dX^{1})\Big(X^{2}(2\pi) - X^{2}(0)\Big)\Big],
\end{align*}
\begin{align*}
b^{j_1 j_2}_k &= \frac{1}{2\pi(2N+1)}\Big[\sum^{N}_{s=1} \Big\{\hat{a}_{s}(dX^{1})\hat{b}_{k-s}(dX^{2}) + \hat{b}_{s}(dX^{1})\hat{a}_{k-s}(dX^{2}) + \hat{a}_{s}(dX^{1})\hat{b}_{k+s}(dX^{2})\\
&\quad - \hat{b}_{s}(dX^{1})\hat{a}_{k+s}(dX^{2})
 + \hat{a}_{s}(dX^{2})\hat{b}_{k-s}(dX^{1}) + \hat{b}_{s}(dX^{2})\hat{a}_{k-s}(dX^{1}) + \hat{a}_{s}(dX^{2})\hat{b}_{k+s}(dX^{1})\\
&\quad - \hat{b}_{s}(dX^{2})\hat{a}_{k+s}(dX^{1})\Big\} + \hat{b}_{k}(dX^{2})\Big(X^{1}(2\pi) - X^{1}(0)\Big) + \hat{b}_{k}(dX^{1})\Big(X^{2}(2\pi) - X^{2}(0)\Big)\Big],
\end{align*}
and
\begin{align*}
\al_{0}(N, j_1, j_2)
&= \frac{1}{2\pi(2N+1)}\Big[\sum^{N}_{s=1} 2\Big\{\hat{a}_{s}(dX^{j_1})\hat{a}_{s}(dX^{j_2}) + \hat{b}_{s}(dX^{j_1})\hat{b}_{s}(dX^{j_2})\Big\}\\
&\hskip 4cm + \Big(X^{j_2}(2\pi) - X^{j_2}(0)\Big)\Big(X^{j_1}(2\pi) - X^{j_1}(0)\Big)\Big],
\end{align*}
and $\hat{a}, \hat{b}$ are defined as before.

\begin{remark}
Each matrix $\Sigma^{j_1 j_2}_{n, N}(t)$ is symmetric, but not necessary positive definite.
\end{remark}

%%%%%%%%%%%%%%%%%%%%%%%%%%%%%%%%%%%%%%%%%%%%%%%%%%
%%%%%%%%%%%%%%%%%%%%%%%%%%%%%%%%%%%%%%%%%%%%%%%%%%
\subsection{Limit theorem}
Since for each $t$, $\Sigma(t)$ is a symmetric non-negative definite matrix, we denote its eigenvalues by $\lambda_i(t), i = 1,\ldots, n,$ such that $\lambda_1(t) \geq \lambda_2(t) \geq \ldots \geq \lambda_d(t) \geq 0$. We also denote by $\hat{\lambda}^{n}_1(t) \geq \hat{\lambda}^{n}_2(t) \geq \ldots \geq \hat{\lambda}^{n}_d(t)$ the eigenvalues of the symmetric matrix $\Sigma^{j_1 j_2}_{n, N}(t)$ defined by either the first or the second symmetrization.

Now we are in a position to state the first main result of this paper.
\begin{thm} \label{theoremFS}
Assume that $\Sigma(t)$ is continuous and for $i = 1,\ldots, d, j= 1,\ldots, d_1$, and $\frac{N}{n} \to 0$ as $n \to \infty,$
$$\bE\Big( \int_0^{2\pi} \big[ \| \bA_i(t) \|^2 + \|\bB_{i,j}(t)\|^4\big] dt \Big) < \infty.$$
Then the following convergence in probability holds
\begin{align*}
\lim_{n, N \to \infty} \sup_{0 \leq t \leq 2\pi} \sum_{i=1}^d | \hat{\lambda}^{n}_i(t) - \lambda_i(t)| = 0.
\end{align*}
\end{thm}

\begin{remark}
This method has been used by Malliavin et al. in \cite{MMR2007} to estimate the eigenvalues of the covariance matrix of a time series of Euro swap rates and Euribor rates. However, these authors did not provide any discussion on the asymptotic behaviour of the estimators. 
\end{remark}

%%%%%%%%%%%%%%%%%%%%%%%%%%%%%%%%%%%%%%%%%%%%%%%%%%
%%%%%%%%%%%%%%%%%%%%%%%%%%%%%%%%%%%%%%%%%%%%%%%%%%
\section{Quadratic Variation  method}
We briefly recall the Quadratic Variation method which was proposed in Ogawa and Wakayama \cite{OW2007}. Let $(h_n)$ be a sequence of positive numbers satisfying $\lim_{n \to \infty} h_n = 0$. For each $t \in (0, T), \ 1 \leq u, v \leq d$, we denote
\begin{align*}
\tilde{\Sigma}_{u,v}^n(t) = \frac{1}{2h_n} \sum_{i:( t-h_n) \leq t_i < t_{i+1} \leq (t+h_n)} \! \! \! (X_u(t_{i+1}) - X_u(t_i)) \times (X_v(t_{i+1}) - X_v(t_i)).
\end{align*}
We suppose that the diffusion coefficient $\bB$ satisfies the following H\"older continuous condition

$H(\alpha)$: For some $\alpha \in (0,1]$, there exists a constant $K$ such that for all $s, t \in [0,T]$,
\begin{equation} \label{Holder}
\bE\|\bB(s) - \bB(t)\|^2 \leq K|s-t|^{2\alpha}.
\end{equation}
For each $t\in (0,T)$, the approximating matrix $\tilde{\Sigma}^n(t)$ is symmetric, non-negative defined. Hence all of its eigenvalues are non-negative.
Let  $\tilde{\lambda}^{n}_1(t) \geq \tilde{\lambda}^{n}_2(t) \geq \ldots \geq \tilde{\lambda}^{n}_d(t)$ denote the eigenvalues of $\tilde{\Sigma}^{N,n}(t)$. Here is the second main result of this paper.
 
\begin{thm} \label{theoremQV1}
Assume that assumption $H(\alpha)$ holds for some $\alpha \in (0,1]$ and $\sup_{t\in (0,T)} \bE (\|\bA(t)\|^4 + \|\bB(t)\|^4)< \infty$. Then we have
$$ \sup_{t \in (h_n, T-h_n)}\sum_{i=1}^d \bE| \tilde{\lambda}^{n}_i(t) - \lambda_i(t)| \leq M\Big( h_n^{\alpha} + \sqrt{\frac{T}{nh_n}}\Big),$$
for some constant $M$ which does not depend on $n$.\\
In particular, if $h_n = O(n^{-1/(2\alpha+1)})$ then
\begin{equation*}
\sup_{t \in (h_n, T-h_n)} n^{\frac{\alpha}{2\alpha+1}} \sum_{i=1}^d \bE| \tilde{\lambda}^{n}_i(t) - \lambda_i(t)| \leq M .
\end{equation*}
\end{thm}

In the following, we will study the case where the price process $X$ contains jump components. More precisely, we suppose that 
 $X$ is a $d$-dimensional stochastic process defined by 
\begin{equation} \label{eqXwJ}
d X (t) = \bA (t, w)dt + \bB (t,w)dW(t) + dJ(t), \ 0 \leq  t \leq T, 
\end{equation}
where $W, \bA, \bB$ are defined as in Section 1 and $J$ is a $d$-dimensional L\'evy process which may depend on $W$. The Blumenthal-Getoor index $\beta$ of $J$ is defined by
$$\beta = \inf \{p \geq  0: \int_{|x| \leq 1} \|x\|^p \nu(dx) < \infty\},$$
where $\nu$ is L\'evy measure of $J$. It is well-known that $\beta \in [0,2]$. 

For each $t \in (0, T), \ 1 \leq u, v \leq d$, we denote
\begin{align*}
\bar{\Sigma}_{u,v}^n(t) = \frac{\pi}{8h_n} &\sum_i \Big ( |\Dt_i(X_u + X_v) \ \Dt_{i+1}(X_u + X_v)| - |\Dt_i X_u \ \Dt_{i+1}X_u| -  |\Dt_i X_v\  \Dt_{i+1}X_v| \Big),
\end{align*}
where the summation is taken over all indices $i$ such that $( t-h_n) \leq t_{i-1} < t_{i+1} \leq (t+h_n)$ and $\Dt_i X_* = X_*(t_i) - X_*(t_{i-1}).$
 
Let $\bar{\lambda}^{n}_1(t) \geq \bar{\lambda}^{n}_2(t) \geq \ldots \geq \bar{\lambda}^{n}_d(t) \geq 0$ denote the eigenvalues of  $\bar{\Sigma}^{n}(t)$. We have the following limit theorem.
 
\begin{thm} \label{theoremQV2}
 Assume that 
\begin{itemize}
\item $H(\alpha)$ holds for some $\alpha \in (0,1],$
\item $\forall q >0, \ \sup_{t\in (0,T)} \bE \|\bA (t)\|^q + \bE \|\bB (t)\|^q < \infty$,
\item $\beta < 2$ and
$\int_{\|x\| \geq 1} \|x\|^2 \nu(dx) < \infty$.
\end{itemize}
Then for any $\gamma \in (0, \frac{\alpha}{2\alpha + 1} \wedge \frac{2-\beta}{2\beta})$, there exists a constant $M$ such that 
 $$ \sup_n \sup_{t \in (h_n, T-h_n)} n^\gamma \sum_{i=1}^d \bE| \bar{\lambda}^{n}_i(t) - \lambda_i(t)| \leq M,$$
 provided that  $h_n = O(n^{-2\gamma})$.
\end{thm}

\begin{remark}
In \cite{ON2010}, the authors introduce another cross volatility estimation scheme for jump diffusion processes by using a threshold parameter to reduce the effect of large size jumps. Furthermore, one can combine the threshold method with the bi-power method presented above to produce a more stable estimation (see \cite{N2010}).
\end{remark}

\begin{remark}    
By following a similar argument as above, one can construct estimation schemes for eigenvalues of the cross volatility matrix of processes which are contaminated by microstructure noise (see \cite{NO2009, OS2011} for some classes of real-time schemes for the estimation of volatility in the noisy case with/without jumps). 
\end{remark}

%%%%%%%%%%%%%%%%%%%%%%%%%%%%%%%%%%%%%%%%%%%%%%%%%%
%%%%%%%%%%%%%%%%%%%%%%%%%%%%%%%%%%%%%%%%%%%%%%%%%%
\section{Numerical Study}
\subsection{Complexity}
The computational cost of the Quadratic Variation method is much less than that of Fourier Series method. Indeed, the cost of computing of the Quadratic Variation method is of order $n^{\frac{2\alpha}{2\alpha+1}}$ while one of Fourier Series method is $N^2n$.

\subsection{Dummy data}
 We consider a stochastic volatility model of Heston's type defined by
\begin{equation}
 \begin{cases}
dX_i(t) = \gamma_i dt + \sum_{j=1}^d \lambda_{ij}\sqrt{v_j(t)}dW_j(t) \\
 dv_j(t) = \alpha_j(b_j - v_j(t))dt + \sigma_j\sqrt{v_j(t)}dB_j(t)
\end{cases}
\label{SVmodel}
\end{equation}
 for $ 1\leq i \leq d, \ \ 1 \leq j \leq d_1, \  t \in [0,T]$, where $\gamma_i, \alpha_j, b_j, \sigma_j, \lambda_{ij}, \ 1\leq i \leq d, \ 1\leq j \leq d_1,$ are constants, $\alpha_j, b_j , \ 1\leq j \leq d_1,$ are positive; $W_j, B_j, 1\leq j \leq d_1$ are mutually independent standard Brownian motions. 

\begin{remark}The class of square-root diffusions 
\begin{equation} \label{squareroot}
 dv(t) = \alpha (b -v(t))dt + \sigma \sqrt{v(t)}dB(t), \quad v(0) = v_0,
\end{equation}
with $W$ a standard one dimensional Brownian motion, was studied in \cite{F1951}. The author showed that if parameters $\alpha, b, v_0$ are positive, then $v(t)$ will stay positive. And if one supposes further that $2\alpha b \geq \sigma^2$, then $v(t)$ is strictly positive for all $t$ with probability $1$. 
\end{remark}

Provided that processes $v$'s can be simulated at discretized time-point $t_k = k\Delta = kT/N_0, \ k= 0,\ldots, N_0,$ one can simulate $X$ by using a simple Euler - Maruyama's scheme as follows
$$X_i(t_{k+1}) = X_i(t_k) + \gamma_i \Delta  + \sqrt{\Delta} \sum_{j=1}^d \lambda_{ij}\sqrt{v_j(t_k)}Z_{jk},$$
where $Z$'s are independent standard normal distribution random variables. 

The simulation of $v$'s is more involved because the values of $v_j(t_k)$ produced by Euler - Maruyama discretization may become negative. We will simulate $v$'s by sampling from the exact transition laws of the processes (see \cite{G2004}).

In the following, we choose $d =5, \ d_1 = 3, \ \gamma_i = b_j = v_j(0) =  i/100, X_i(0) = 1,  \ \alpha_j = 2, \lambda_{ij} = (-1)^{i+j}\sin(ij), \ 1\leq i \leq d, \ 1\leq j \leq d_1.$ We choose $\sigma_j = \sqrt{2b_j \alpha_j}$ and $T = 2\pi$ to make simulation easier.

Based on the sample data of $X$, we use both the Quadratic Variation and Fourier Series methods, as stated in the previous sections, to estimate the cross volatility matrices of $X$ as a function of time and after that we calculate the eigenvalues of each estimated matrix. In particular, since the volatility coefficients of $X$ satisfy assumption (\ref{Holder}) with $\alpha = 1/2$, we choose $h_n = TN_0^{-1/2}$ for the Quadratic Variation scheme. Besides, for the Fourier Series method, we calculate the Fourier coefficients of the cross volatilities up to the Nyquist frequency $2N = N_0/2$ (see \cite{P1981}).

We observe the mean square pathwise errors $MSE$ and $mSE$ defined as follows: Suppose that for each $k=0,\ldots, N_0$, $\check{\Sigma}(t_k)$ is an estimator of matrix $\Sigma(t_k)$. We denote by $\check{\lambda}_1(t_k)$ and $\check{\lambda}_d(t_k)$ the maximum and  minimum eigenvalues of $\check{\Sigma}(t_k)$. We also denote by $\lambda_1(t_k)$ and $\lambda_d(t_k)$ the maximum and  minimum eigenvalues of $\Sigma(t_k)$.  Then we measure the errors of the estimations on the whole paths by
$$MSE(\Sigma, \check{\Sigma}) = \frac{1}{N_0}\sum_{k=1}^{N_0}|\check{\lambda}_1(t_k) - \lambda_1(t_k)|^2,$$
and
$$mSE(\Sigma, \check{\Sigma}) = \frac{1}{N_0}\sum_{k=1}^{N_0}|\check{\lambda}_d(t_k) - \lambda_d(t_k)|^2.$$

%%%%%%%%%%%%%%%%%%%%%%%%%%%%%%%%%%%%%%%%%%%%%%%%%%
%%%%%%%%%%%%%%%%%%%%%%%%%%%%%%%%%%%%%%%%%%%%%%%%%%
\subsubsection{The results of the first Fourier Series method}
\begin{figure} % figuur 1
\begin{center}
\hskip -1cm 
\includegraphics[width=6.5cm,height=4.5cm]{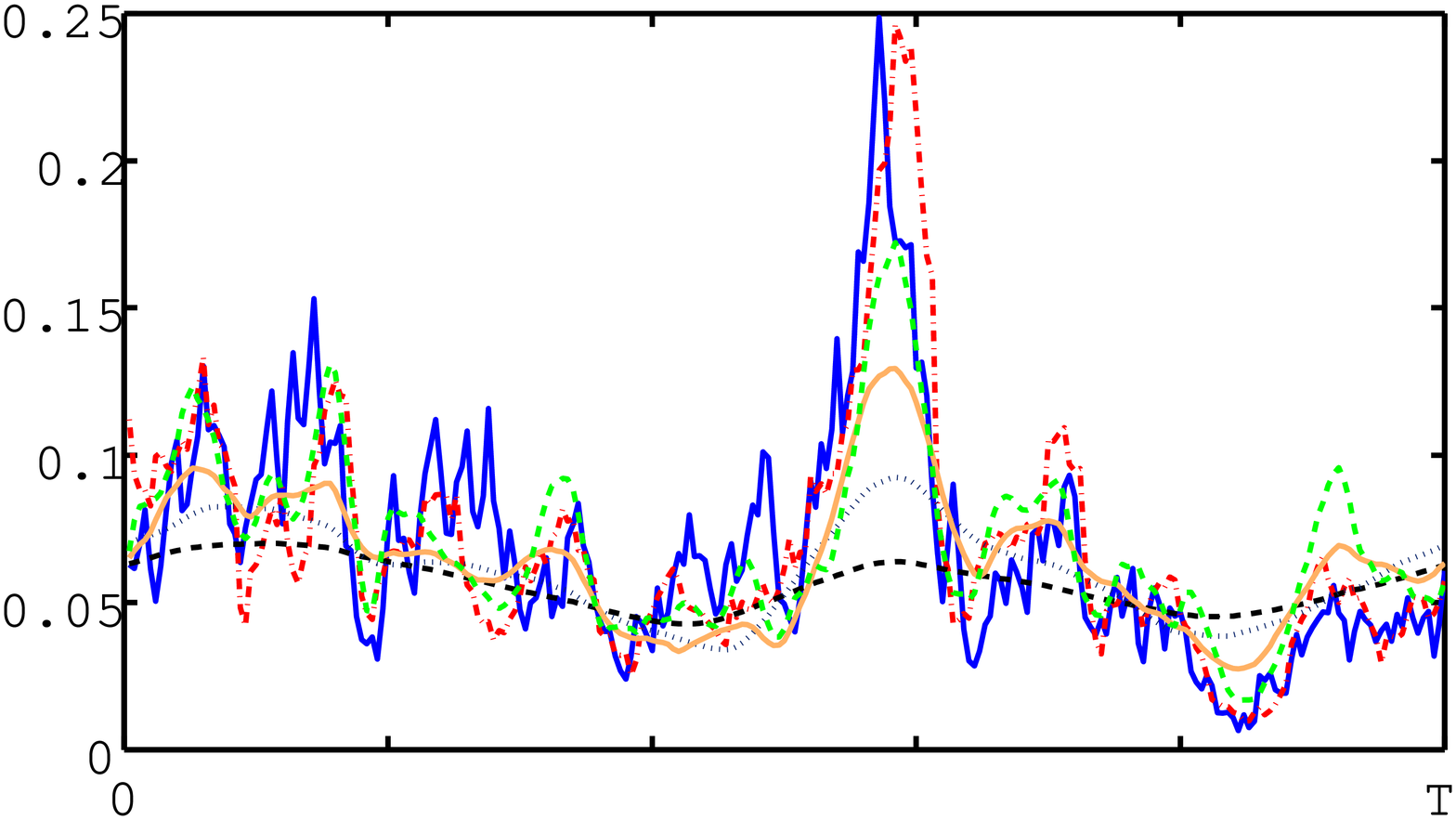}
\includegraphics[width=7.5cm,height=4.5cm]{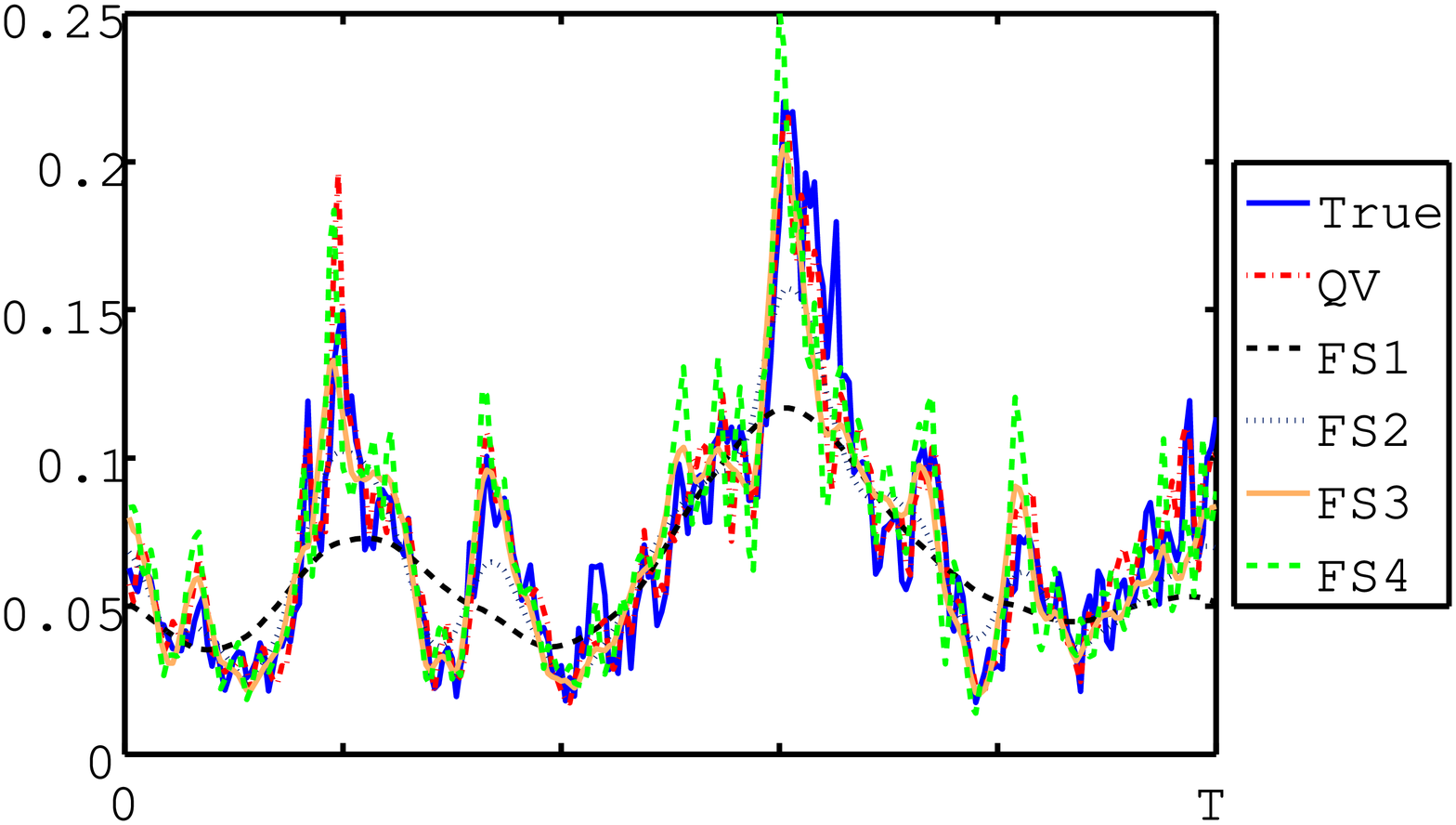}
\end{center}
\vskip -1cm
\caption{Maximum eigenvalue correspond to the Quadratic Variation method and the first Fourier Series method (left:$N_0 = 10^3$, right:$N_0 = 10^4$)}
\label{f3_1}
\end{figure}

\begin{figure} % figuur 1
\begin{center}
\hskip -1cm 
 \includegraphics[width=6.5cm,height=4.5cm]{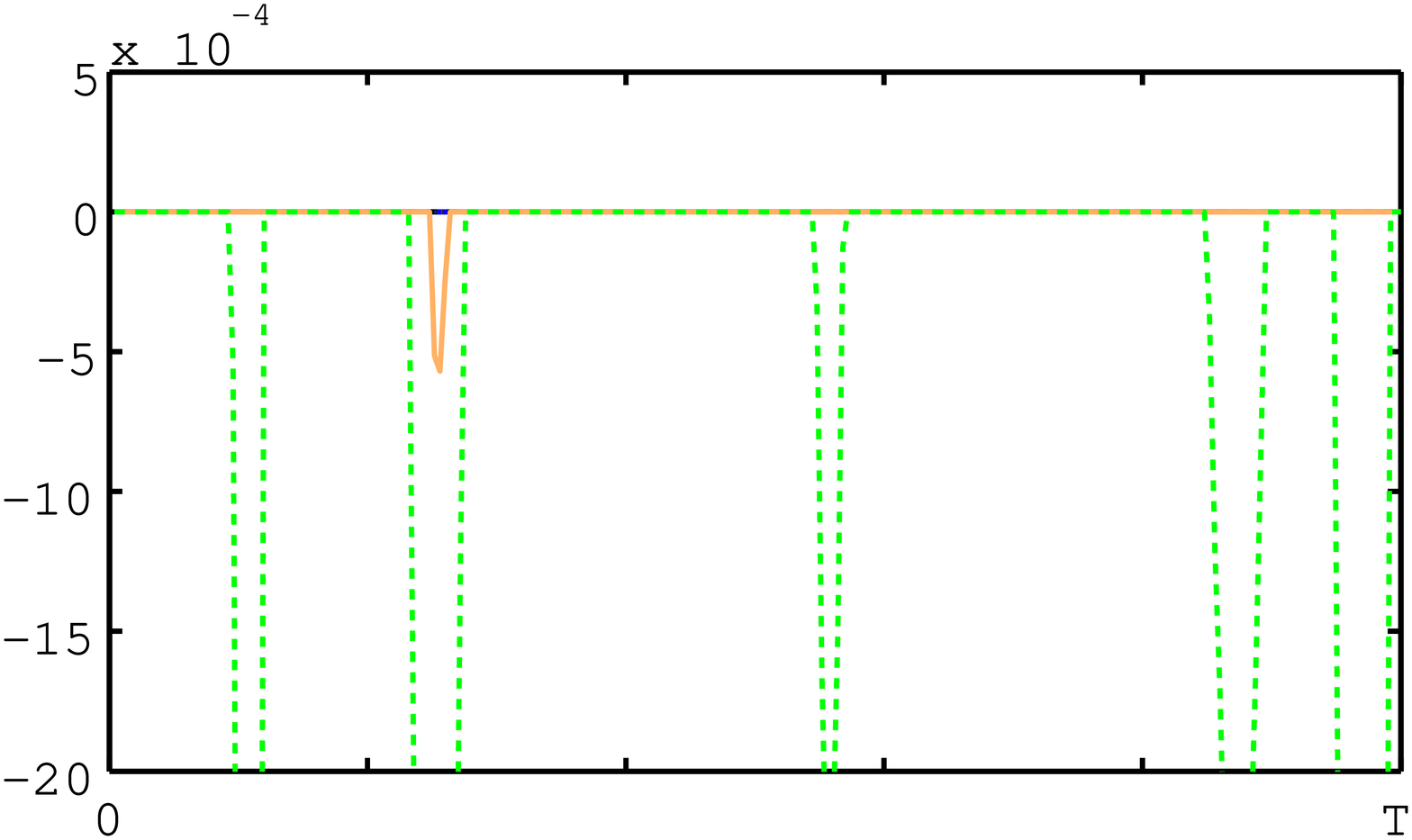}
 \includegraphics[width=7.5cm,height=4.5cm]{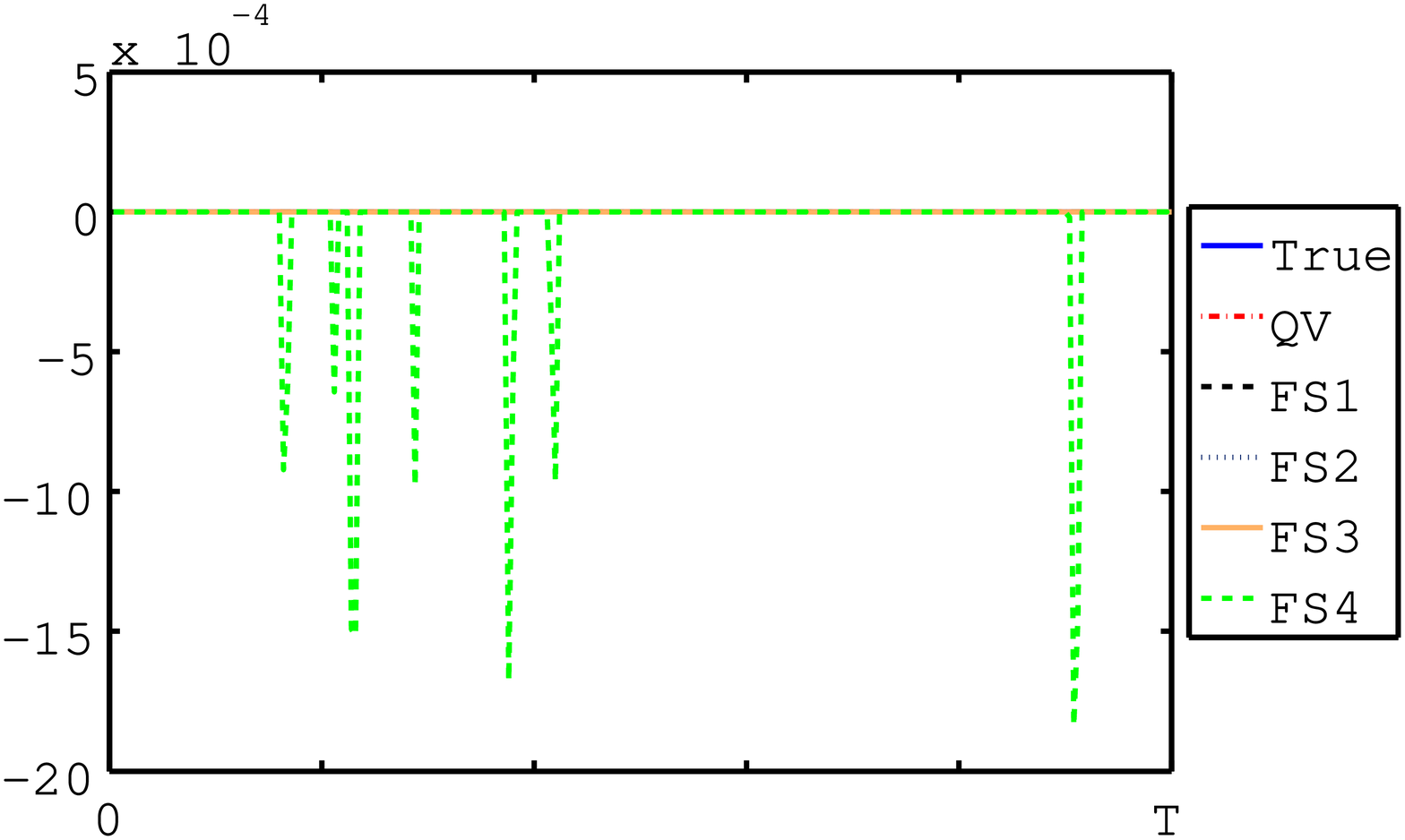}
\end{center}
\vskip -1cm
\caption{Minimum eigenvalue correspond to the Quadratic Variation method and the first Fourier Series method (left:$N_0 = 10^3$, right:$N_0 = 10^4$)}
\label{f4_1}
\end{figure}

The simulations show that the Fourier Series estimate does not work well near $0$ and $T$. In order to have a better understanding of errors of each estimation methods at "normal" time, we eliminate $10$ percent of the estimated cross volatilities near the two end points $0$ and $T$ when we calculate the mean square pathwise errors for each symmetrization of the Fourier Series method and the Quadratic Variation method. The means of $mSE$ and $MSE$ of each method are showed in Table \ref{errors1} (Note that in all tables we use $\epsilon$ for value less than $10^{-10}$). Here QV and FS stand for Quadratic Variation and Fourier Series methods, respectively. FS$i$, $i = 1, 2, 3, 4,$ stand for Fourier Series estimation using smooth kernel with $\delta = TN_0^{-0.1(i+2)}$, respectively. Figures \ref{f3_1} and \ref{f4_1} show the estimations of $\lambda_M$ and $\lambda_m$ during $(0, T)$ with $N_0= 10^3$ and $N_0 = 10^4$.

\begin{table}
\begin{center}
\begin{tabular}{ l r r r r r r r r }
& $N_0$ & QV & FS & FS$1$ & FS$2$ & FS$3$ & FS$4$\\
 \hline 
MSE  & $10^2$ & 23 & 100 & 21 & 21 & 23 & 31\\
mSE &  & $\epsilon$  & 6.882 & $\epsilon$ & 0.006 & 0.071 & 0.391 \\ 
 \hline 
MSE  & $10^3$ & 7 & 88 & 15 & 11 & 9 & 12\\
mSE &  & $\epsilon$ & 7.889 & $\epsilon$ & $\epsilon$ & 0.01 & 0.077 \\
  \hline 
MSE & $10^4$ & 2 & 93 & 9 & 5 & 4 & 6 \\
mSE &  & $\epsilon$ & 7.609 & $\epsilon$ & $\epsilon$ & $\epsilon$ & 0.007 \\
  \hline  
\end{tabular}
\caption{Means of $MSE$ and $mSE$ ($\times 10^{-4}$) correspond to the Quadratic Variation method and the first Fourier Series method }\label{errors1}
\end{center}
\end{table}

Remark that we remove the graph of FS since it oscillates violently making the whole picture difficult to see. The Fourier Series scheme using the modified F\'ejer kernel is able to produce a good estimate provided that one can choose a correct value for the parameter $\delta$. However, Table \ref{errors1} together with  Figure \ref{f3_1} shows that this estimation is very sensitive to the choice of $\delta$. And to the best of our knowledge, there is still no effective way to select a good $\delta$.

Another disadvantage of the Fourier Series method is evident from Figure \ref{f4_1}. One can see that FS3 and FS4 schemes may produce a negative estimated values of eigenvalues of the cross volatility matrix at a significant level. This drawback happens because the estimated cross volatility matrices using Fourier Series method may be not non-negative definite in general.
\subsubsection{The results of the second Fourier Series method}
\begin{figure} % figuur 1
\begin{center}
\hskip -1cm 
\includegraphics[width=6.5cm,height=4.5cm]{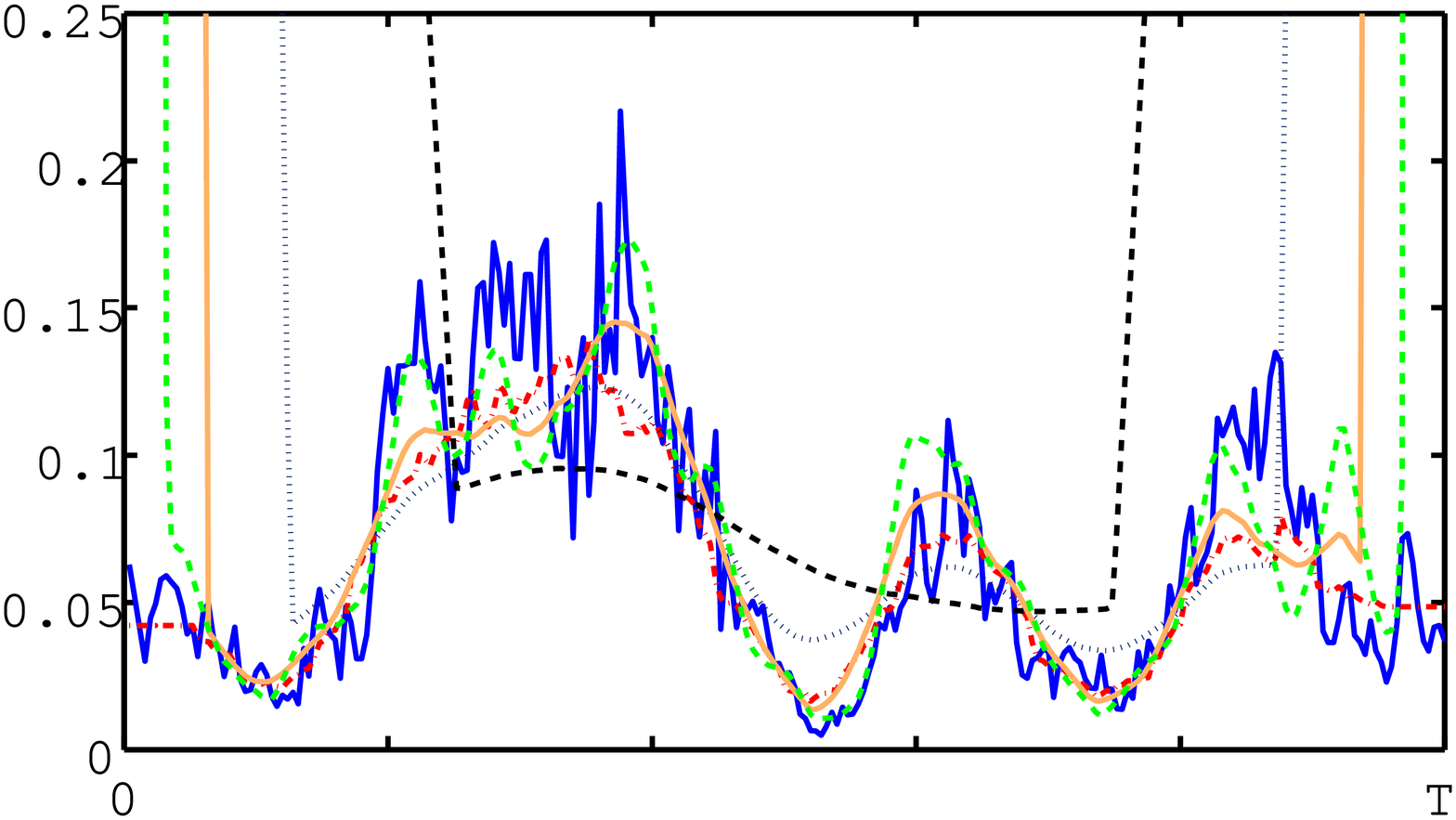}
\includegraphics[width=7.5cm,height=4.5cm]{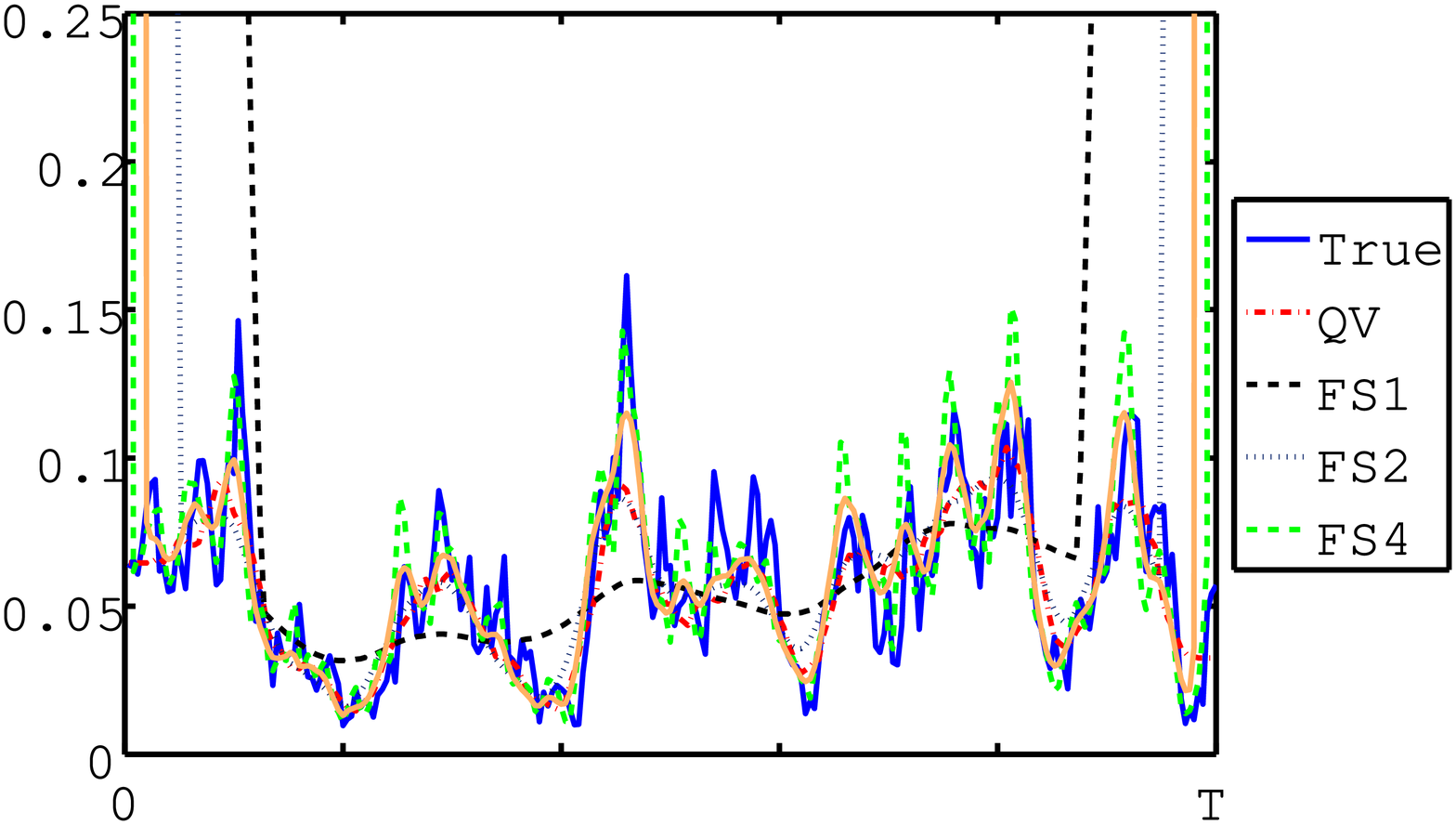}
\end{center}
\vskip -1cm
\caption{Maximum eigenvalue correspond to the Quadratic Variation method and the second Fourier Series method with first symmetrization (left:$N_0 = 10^3$, right:$N_0 = 10^4$)}
\label{fs1_3}
\end{figure}

\begin{figure} % figuur 1
\begin{center}
\hskip -1cm 
\includegraphics[width=6.5cm,height=4.5cm]{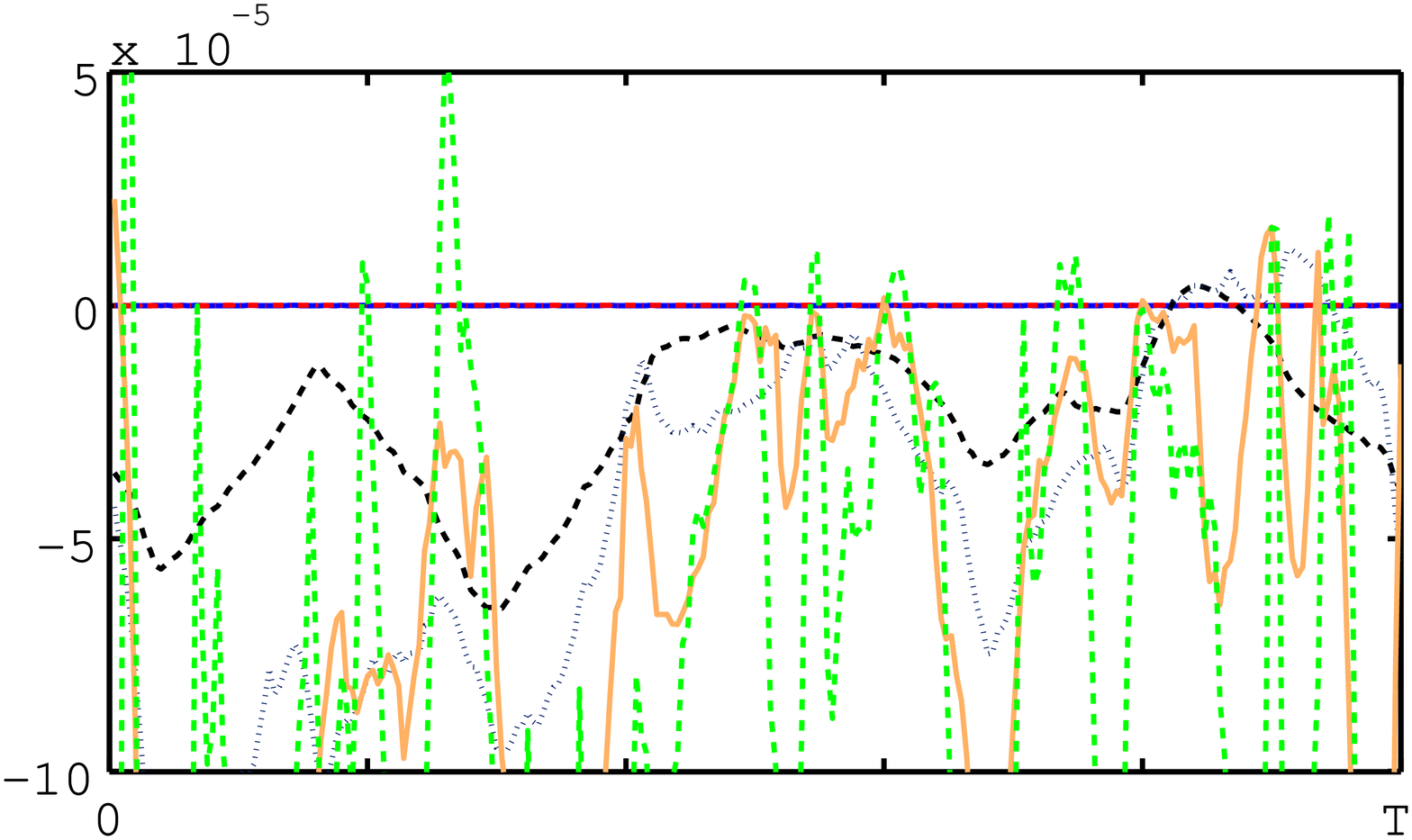}
\includegraphics[width=7.5cm,height=4.5cm]{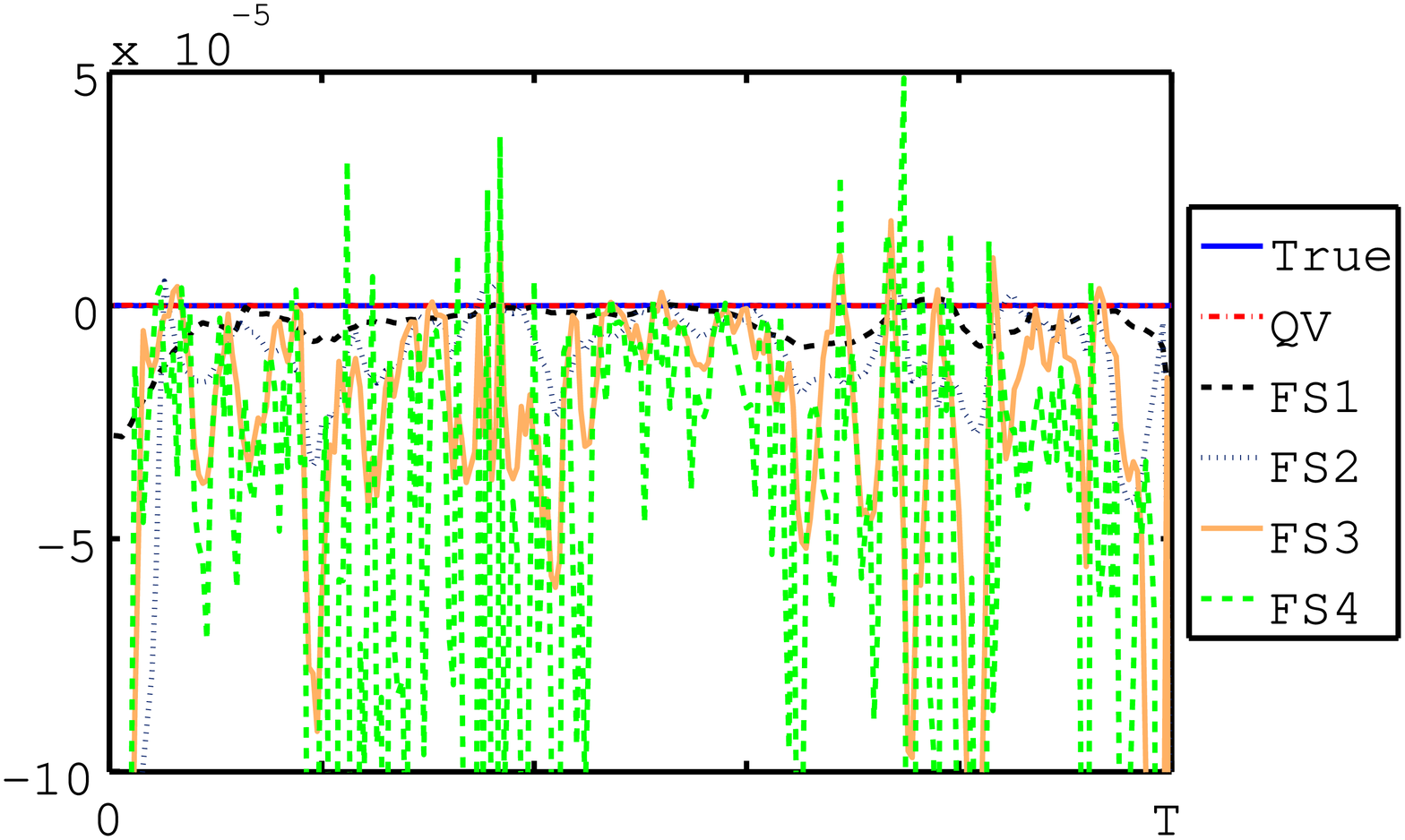}
\end{center}
\vskip -1cm
\caption{Minimum eigenvalue correspond to the Quadratic Variation method and the second Fourier Series method with first symmetrization (left:$N_0 = 10^3$, right:$N_0 = 10^4$)}
\label{fs1_4}
\end{figure}

\begin{figure} % figuur 1
\begin{center}
\hskip -1cm 
\includegraphics[width=6.5cm,height=4.5cm]{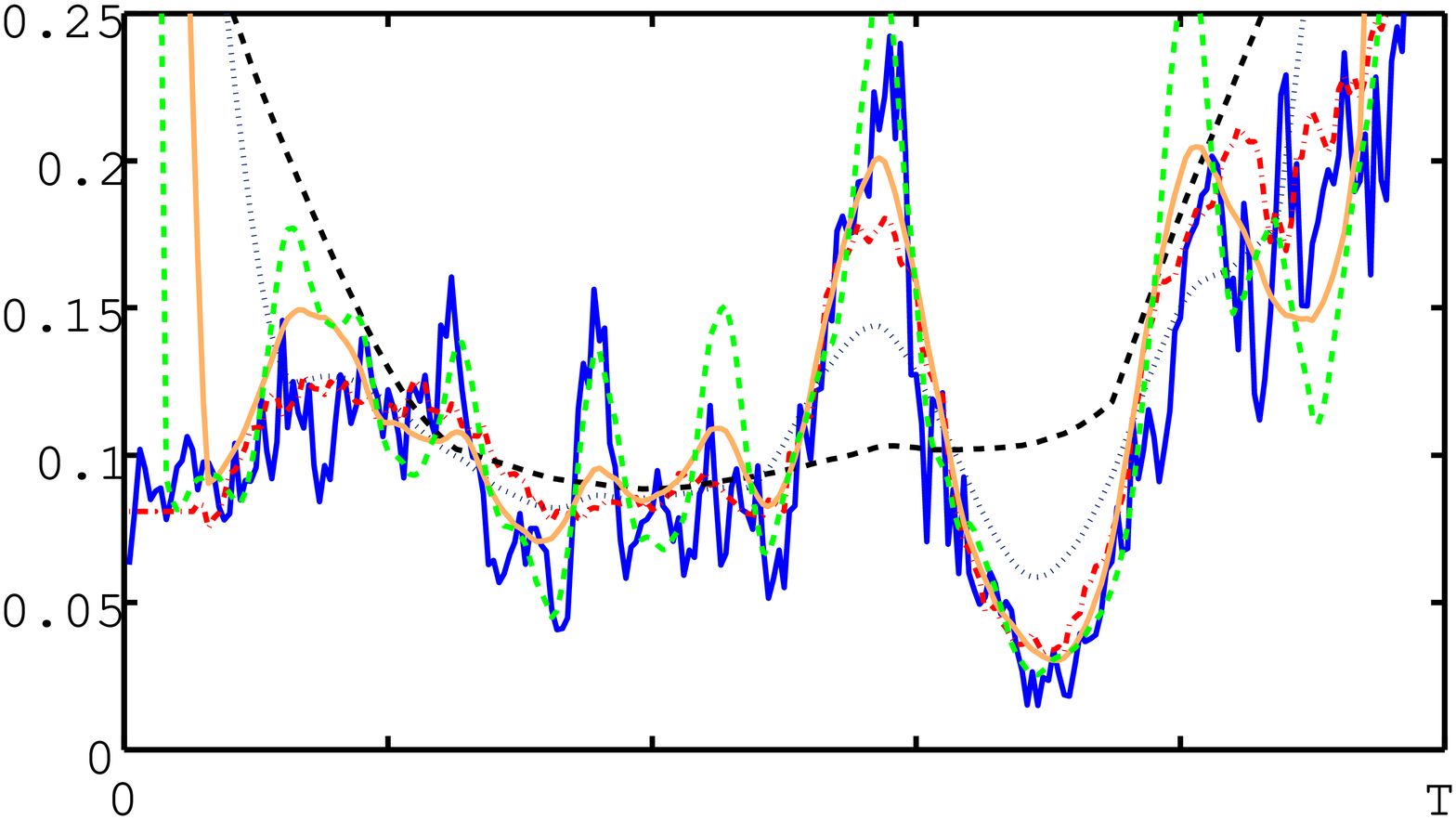}
\includegraphics[width=7.5cm,height=4.5cm]{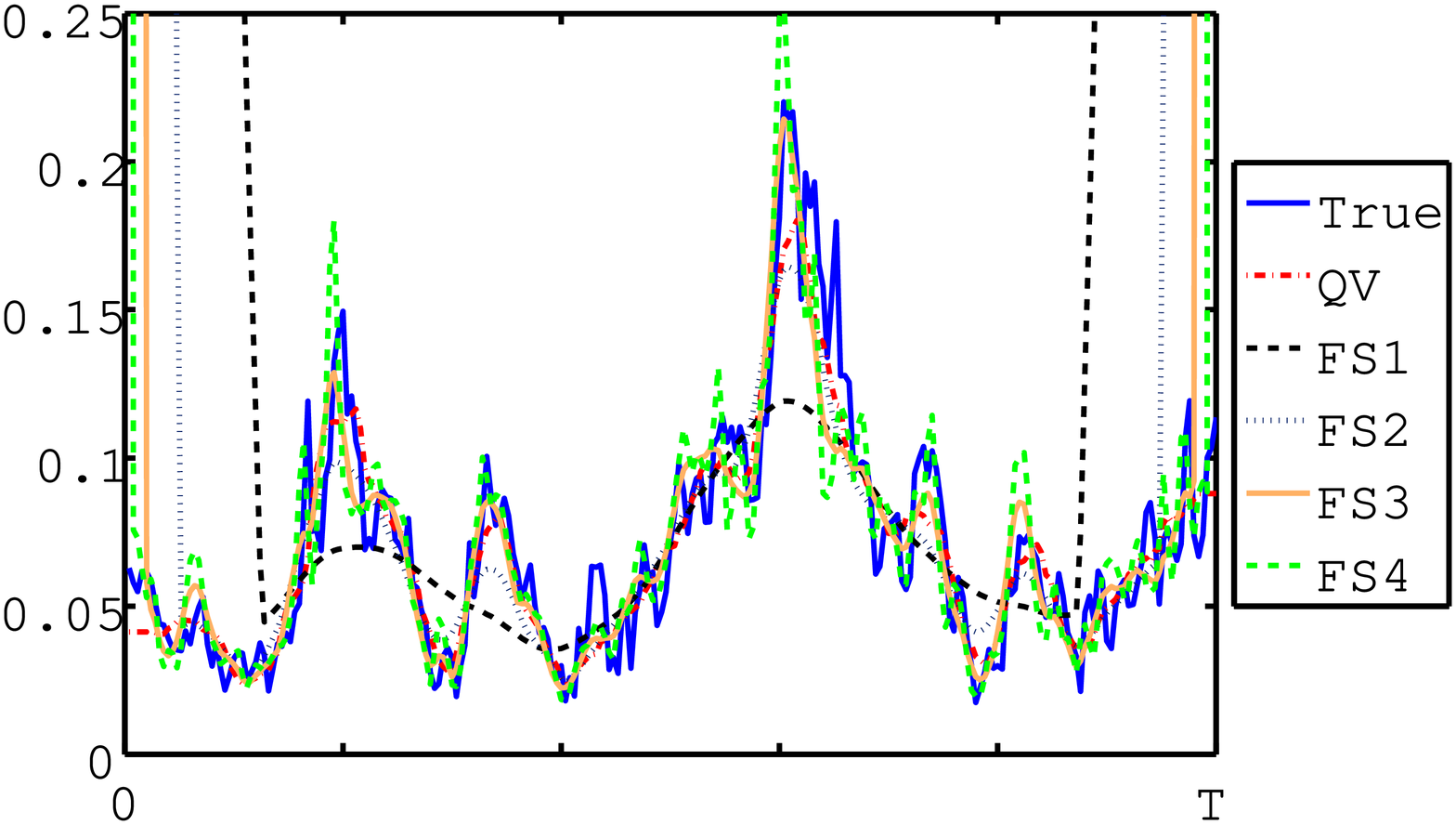}
\end{center}
\vskip -1cm
\caption{Maximum eigenvalue correspond to the Quadratic Variation method and the second Fourier Series method with second symmetrization (left:$N_0 = 10^3$, right:$N_0 = 10^4$)}
\label{fs2_3}
\end{figure}

\begin{figure} % figuur 1
\begin{center}
\hskip -1cm 
\includegraphics[width=6.5cm,height=4.5cm]{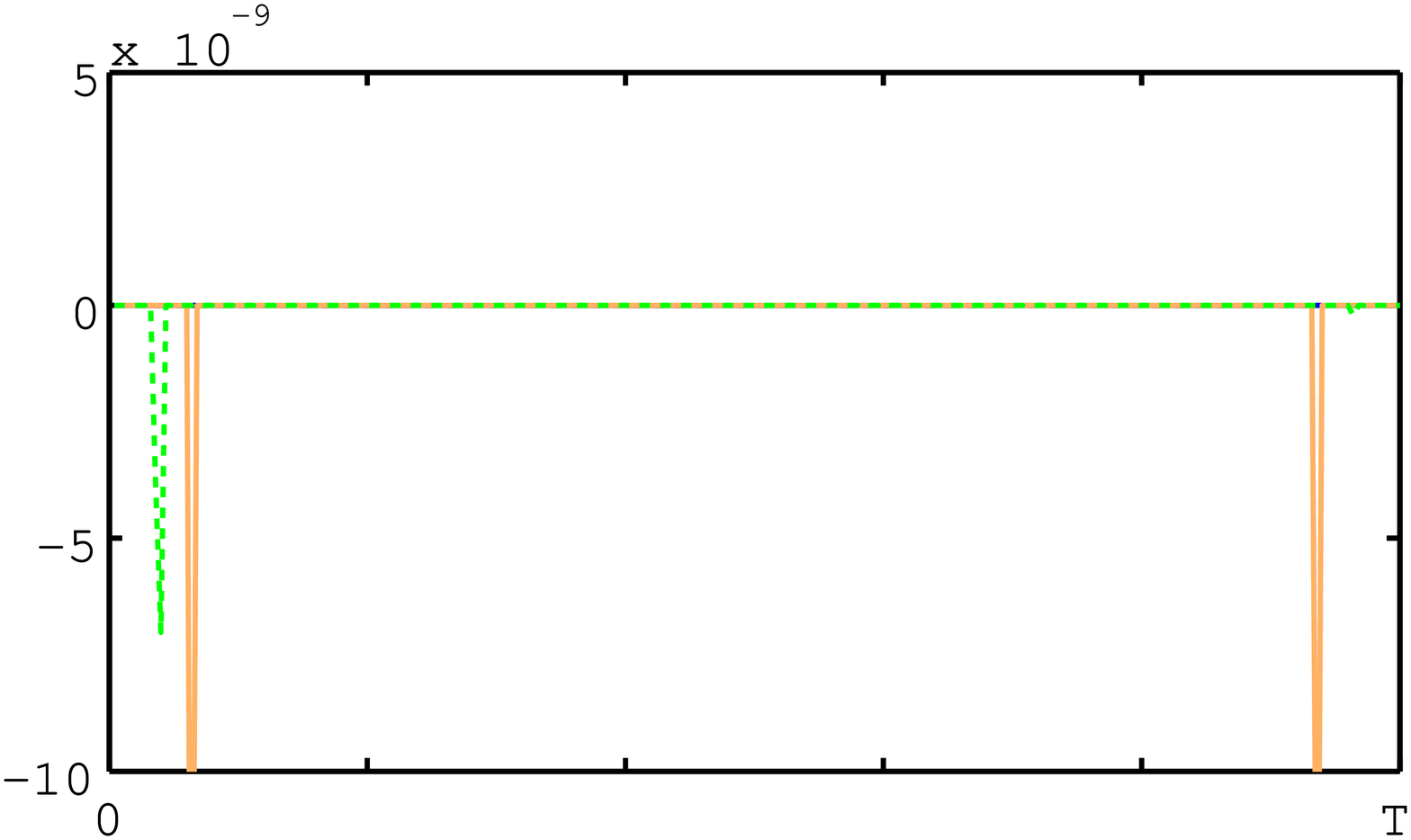}
\includegraphics[width=7.5cm,height=4.5cm]{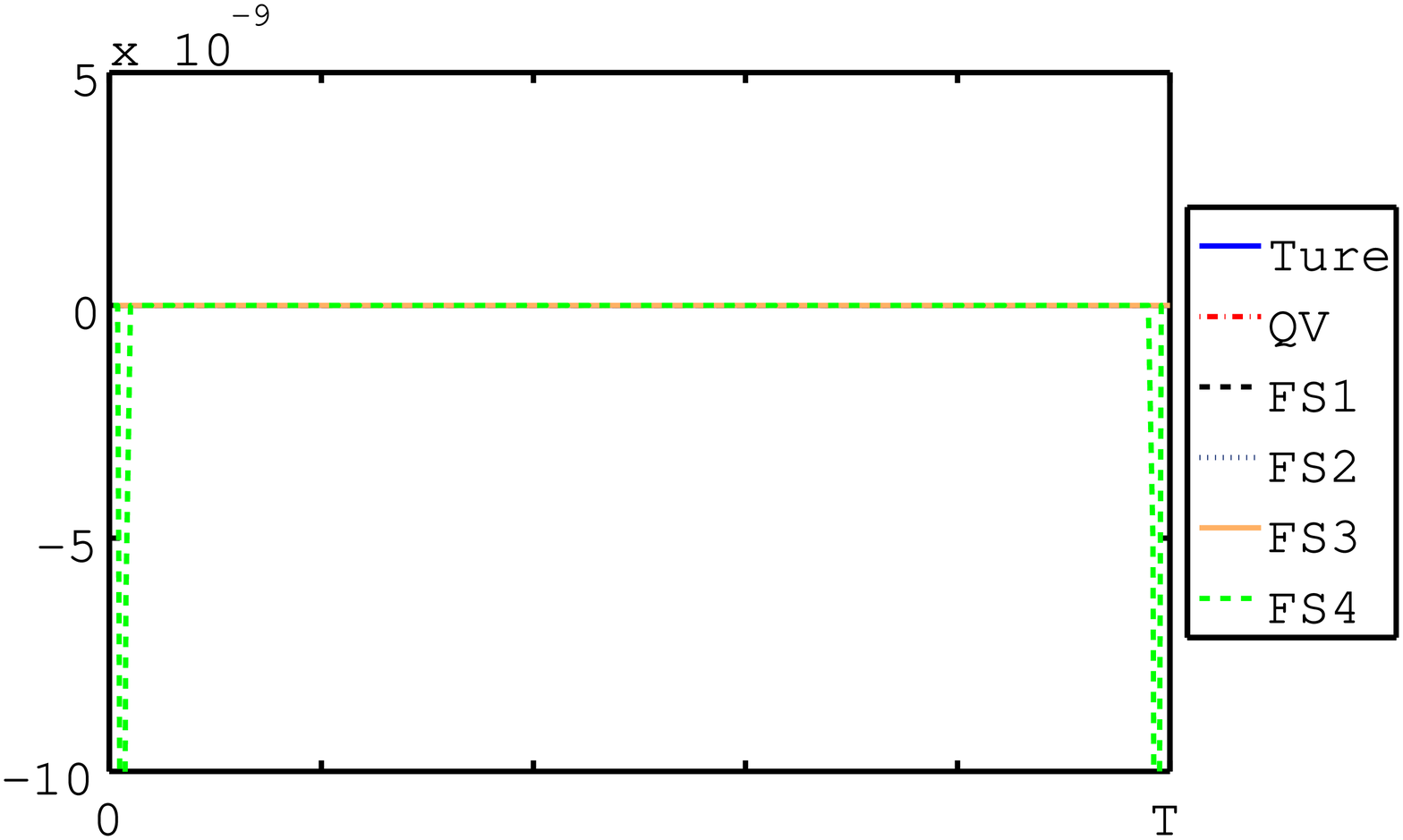}
\end{center}
\vskip -1cm
\caption{Minimum eigenvalue correspond to the Quadratic Variation method and the second Fourier Series method with second symmetrization (left:$N_0 = 10^3$, right:$N_0 = 10^4$)}
\label{fs2_4}
\end{figure}

We use the same notations as above.
Table \ref{errors2_s1} and Table \ref{errors2_s2} show the means of $mSE$ and $MSE$ of each method while Fourier Series method modified by first symmetrization and second symmetrization, respectively.
Figures \ref{fs1_3} and \ref{fs1_4} show the estimations of $\lambda_M$ and $\lambda_m$ during $(0, T)$ with $N_0= 10^3$ and $N_0 = 10^4$ with first symmetrization, and Figures \ref{fs2_3} and \ref{fs2_4} show the estimations of $\lambda_M$ and $\lambda_m$ during $(0, T)$ with $N_0= 10^3$ and $N_0 = 10^4$ with second symmetrization.

Base on two symmetrization methods, FS3 and FS4 schemes give us a good result which is shown in Figures \ref{fs1_3} and \ref{fs2_3}.
Although the Fourier Series estimators are still not non-negative definite, a negative value of the estimate of eigenvalue of the cross volatility matrix is not significant as shown in Figures \ref{fs1_4} and \ref{fs2_4}.

\begin{table}
\begin{center}
\begin{tabular}{ l r r r r r r r r }
& $N_0$ & QV & FS & FS$1$ & FS$2$ & FS$3$ & FS$4$\\
 \hline 
MSE  & $10^2$ & 18 & 399 & 456 & 391 & 164 & 106 \\
mSE &  & $\epsilon$ & 3 & 10 & 32 & 96 & 1721 \\ 
 \hline 
MSE  & $10^3$ & 8 & 298 & 43 & 8 & 10 & 65 \\
mSE &  & $\epsilon$ & $\epsilon$ & $\epsilon$ & $\epsilon$ & $\epsilon$ & $\epsilon$ \\
  \hline 
MSE & $10^4$ & 3 & 31 & 5 & 3 & 5 & 67 \\
mSE &  & $\epsilon$ & $\epsilon$ & $\epsilon$ & $\epsilon$ & $\epsilon$ & $\epsilon$ \\
  \hline  
\end{tabular}
\caption{Means of $MSE$ and $mSE$ ($\times 10^{-4}$) correspond to the Quadratic Variation method and the second Fourier Series method with first symmetrization}\label{errors2_s1}
\end{center}
\end{table}

\begin{table}
\begin{center}
\begin{tabular}{ l r r r r r r r r }
& $N_0$ & QV & FS & FS$1$ & FS$2$ & FS$3$ & FS$4$\\
 \hline 
MSE  & $10^2$ & 20 & 427 & 491 & 424 & 175 & 100 \\
mSE &  & $\epsilon$ & $\epsilon$ & $\epsilon$ & $\epsilon$ & 10 & 3768 \\ 
 \hline 
MSE  & $10^3$ & 8 & 207 & 33 & 9 & 11 & 65 \\
mSE &  & $\epsilon$ & $\epsilon$ & $\epsilon$ & $\epsilon$ & $\epsilon$ & 14 \\
  \hline 
MSE & $10^4$ & 3 & 32 & 4 & 3 & 5 & 66 \\
mSE &  & $\epsilon$ & $\epsilon$ & $\epsilon$ & $\epsilon$ & $\epsilon$ & $\epsilon$ \\
  \hline  
\end{tabular}
\caption{Means of $MSE$ and $mSE$ ($\times 10^{-4}$) correspond to the Quadratic Variation method and the second Fourier Series method with second symmetrization}\label{errors2_s2}
\end{center}
\end{table}

In our simulation, the Quadratic Variation method works quite well. Its mean square pathwise error is strictly less than the ones of Fourier Series method. In addition, because the estimated cross volatility matrices using the Quadratic Variation method are always symmetric and non-negative definite, all of their eigenvalues are non-negative. Finally, the computation time of the Quadratic Variation scheme is less than 1/100 of the Fourier Series scheme. 

%%%%%%%%%%%%%%%%%%%%%%%%%%%%%%%%%%%%%%%%%%%%%%%%%%
%%%%%%%%%%%%%%%%%%%%%%%%%%%%%%%%%%%%%%%%%%%%%%%%%%
\section{Proofs}
In this section, we sketch the proofs of the main results in Sections 2 and 3. First we need the following auxiliary inequality (see \cite{AGZ2010}). 

\begin{lem}[Hoffman-Wielandt]  \label{HWlemma}
Let $A, B$ be $N \times N$ symmetric matrices, with eigenvalues $\lambda^A_1 \leq \lambda^A_2 \leq \ldots \leq \lambda^A_N$ and $\lambda^B_1 \leq \lambda^B_2 \leq \ldots \leq \lambda^B_N$. Then
$$ \sum_{i=1}^N |\lambda^A_i - \lambda^B_i|^2 \leq tr(A-B)^2.$$
\end{lem}

\subsection{Proof of Theorem \ref{theoremFS}}
By using a similar argument as in the proof of Theorem 3.4 (\cite{MM2009}), one can show that the following convergence in probability holds
\begin{align*}
\lim_{n, N \to \infty} \sup_{0 \leq t \leq 2\pi} \| \hat{\Sigma}^{N,n}(t) - \Sigma(t)\| = 0.
\end{align*} 
Hence, one also has 
\begin{align*}
\lim_{n, N \to \infty} \sup_{0 \leq t \leq 2\pi} tr( \hat{\Sigma}^{N,n}(t) - \Sigma(t))^2 = 0 , \ \text{in probability}.
\end{align*}
By applying Lemma \ref{HWlemma}, we get the desired result.

\subsection{Proof of Theorem \ref{theoremQV1}}
After some elementary calculations, one gets from Lemma \ref{HWlemma} that
\begin{align} \label{eqnQV1}
\sum_{i=1}^d | \tilde{\lambda}^{n}_i(t) - \lambda_i(t)| \leq \sqrt{d}\sum_{i,j=1}^d |\tilde{\Sigma}^{n}_{ij}(t) - \Sigma_{ij}(t)|,
\end{align}
 for all $t \in (0,T)$. On the other hand, it follows from Proposition 3.3 (\cite{OW2007}) that there exists a constant $M > 0$ such that for all $n$ and all $t \in (h_n, T-h_n)$, one has
\begin{align*}
\Big( h_n^{\alpha} + \sqrt{\frac{T}{nh_n}}\Big)^{-1} \sum_{i, j=1}^d \bE| \tilde{\Sigma}^{n}_{ij}(t) - \Sigma_{ij}(t)| \leq M,
 \end{align*}
 which concludes Theorem  \ref{theoremQV1}.

\subsection{Proof of Theorem \ref{theoremQV2}}
It is not hard to see from Propositions 3.13 and 3.14 in \cite{NO2009} that for any $\gamma \in (0, \frac{\alpha}{2\alpha + 1} \wedge \frac{2-\beta}{2\beta})$, there exists a constant $M$ such that 
\begin{align*}
n^\gamma \sum_{i, j=1}^d \bE| \bar{\Sigma}^{n}_{ij}(t) - \Sigma_{ij}(t)| \leq M,\end{align*}
for all $n$ and $t \in (0,T)$. This fact together with estimate (\ref{eqnQV1}) yields the desired result.

%%%%%%%%%%%%%%%%%%%%%%%%%%%%%%%%%%%%%%%%%%%%%%%%%%
%%%%%%%%%%%%%%%%%%%%%%%%%%%%%%%%%%%%%%%%%%%%%%%%%%
\section{Conclusions}
In this paper we studied two methods to estimate the eigenvalues of spot cross volatility matrix. The empirical studies show that in comparison with the Fourier Series method, the Quadratic Variation method is easier to implement, is much faster and is able to avoid the negative eigenvalue problem. The Quadratic Variation method is also applicable to diffusion processes with jumps for which the Fourier Series method is unsuitable. 

%%%%%%%%%%%%%%%%%%%%%%%%%%%%%%%%%%%%%%%%%%%%%%%%%%
%%%%%%%%%%%%%%%%%%%%%%%%%%%%%%%%%%%%%%%%%%%%%%%%%%

\end{document}